\documentclass[showpacs,twocolumn]{revtex4}
\usepackage{amsmath}
\usepackage{amssymb}
\usepackage{graphicx}
\usepackage{bm}
\usepackage{dcolumn}
\newcommand{\bpm}{\begin{pmatrix}}
\newcommand{\epm}{\end{pmatrix}}

\begin{document}

\title{Crossover from the weak to strong-field behavior of the longitudinal
interlayer magnetoresistance in quasi-two-dimensional conductors}
\author{A. D. Grigoriev}
\affiliation{Samara State University, Russia}
\author{P. D. Grigoriev}
\email{grigorev@itp.ac.ru}
\affiliation{L.D. Landau Institute for Theoretical Physics, Chernogolovka, Russia}
\date{\today }

\begin{abstract}
We investigate the monotonic growth of longitudinal interlayer
magnetoresistance $\bar{R}_{zz}\left( B_{z}\right) $, analytically and
numerically in the self-consistent Born approximation. We show that in a
weak magnetic field the monotonic part of $\bar{R}_{zz}\left( B_{z}\right) $
is almost constant and starts to grow only above the crossover field $B_{c}$%
, when the Landau levels (LL) become isolated, i.e. when the LL separation
becomes greater than the LL broadening. In higher field $B_{z}\gg B_{c}$, $%
\bar{R}_{zz}\left( B_{z}\right) \propto B_{z}^{1/2}$ in agreement with
previous works.
\end{abstract}

\maketitle

\section{ Introduction}

Magnetoresistance (MR) in strongly anisotropic layered metals is extensively
studied during last decades, because it provides a powerful tool to
determine the electronic properties of various layered materials, including
high-temperature superconductors \cite%
{HusseyNature2003,AbdelNature2006,ProustNature2007,AbdelPRL2007AMRO,McKenzie2007,DVignolle2008,Bangura2008,HelmNd2009,HelmNd2010,BaFeAs2011,Graf2012}%
), organic metals (see, e.g., Refs. \cite%
{MarkReview2004,KartPeschReview,LebedBook} for recent reviews),
heterostructures \cite{Kuraguchi2003} etc. The standard three-dimensional
theory of MR \cite{Abrik,Shoenberg,Ziman,KartPeschReview}, based on the $%
\tau $-approximation, is not valid in the two-dimensional (2D) electron
system because of the high Landau-level (LL) degeneracy (see, e.g., Refs.
\cite{Ando1,Fogler1997} or Refs. \cite{KukushkinUFN1988},\cite%
{DmitrievRMP2012} for review) even in the fields insufficient for the
quantum Hall effect (QHE) \cite{QHE,HuckesteinRMP1995}. In strongly
anisotropic layered quasi-2D metals, when the interlayer transfer integral $%
t_{z}$ is less than the LL separation $\hbar \omega _{c}=\hbar eB/m^{\ast }c$%
, the standard MR theory\cite{Abrik,Shoenberg,Ziman,KartPeschReview} is also
inapplicable. In particular, it predicts only a transverse MR, while the
strong longitudinal interlayer MR $R_{zz}\left( B_{z}\right) $ is observed
in various compounds as a general feature of quasi-2D conductors \cite%
{SO,Coldea,PesotskiiJETP95,Zuo1999,W2,W3,W4,Incoh2009,Kang,WIPRB2012}. In
spite of a considerable theoretical attention to MR in quasi-2D compounds,%
\cite{MosesMcKenzie1999,PhSh,Shub,ChampelMineev,Gvozd2004,Gvozd2007} this
longitudinal interlayer MR has been explained only recently\cite%
{WIPRB2011,WIFNT2011,WIJETP2011} in the high-field limit, when the LLs do
not overlap, i.e. when the LL separation $\hbar \omega _{c}$ is greater than
the LL broadening $2\Gamma _{0}=\hbar /\tau $, while $t_{z}\ll \Gamma _{0}$.
Qualitatively, the longitudinal interlayer MR $R_{zz}\left( B_{z}\right)
\propto B_{z}^{1/2}$ originates from the monotonic growth of the LL width $%
\Gamma \left( B_{z}\right) \propto B_{z}^{1/2}$, well-known in a 2D electron
system.\cite{Ando1} This LL width, being equal to the imaginary part of
electron self-energy $\left\vert \text{Im}\Sigma \right\vert $, enters the
denominator of the interlayer conductivity similarly to the scattering rate.%
\cite{WIPRB2011,WIFNT2011,WIJETP2011} Various LL shapes give slightly
different coefficients $\eta \sim 1$ in the high-field dependence of $%
R_{zz}\left( B_{z}\right) $ \cite{WIJETP2011}:
\begin{equation}
R_{zz}\left( B_{z}\right) /R_{zz}\left( 0\right) =\eta \sqrt{\hbar \omega
_{c}/\Gamma _{0}}.  \label{Rzz0}
\end{equation}%
The Lorenztian LL shape gives $\eta =\sqrt{4/\pi }$, the non-crossing (or
single-site) approximation\cite{Ando2} gives\cite{CommentNC} $\eta =3\sqrt{%
\pi }/8\approx 0.665$, and the same value $\eta =3\sqrt{\pi }/8$ is obtained
in the self-consistent Born approximation (SCBA)\cite{Ando1} [see Eq. (\ref%
{sLim}) below]. In Ref. \cite{GrigPRB2013} the calculation of $R_{zz}\left(
B_{z}\right) $ was generalized to a finite $t_{z}\gtrsim \Gamma _{0}$ but
still in the high-field limit $\hbar \omega _{c}>4t_{z}$. The behavior at $%
\hbar \omega _{c}\lesssim \Gamma _{0}$ is still unknown. The smooth
dependence%
\begin{equation}
R_{zz}\left( B_{z}\right) \propto \left[ \left( \hbar \omega _{c}/\Gamma
_{0}\right) ^{2}+1\right] ^{1/4},  \label{Rzz1}
\end{equation}%
assumed in Refs. \cite{WIPRB2011,WIPRB2012} to compare with experimental
data, does not have a theoretical substantiation. The aim of this paper is
to calculate the longitudinal interlayer magnetoresistance at $t_{z}\ll
\hbar \omega _{c}\lesssim \Gamma _{0}$.

\section{Analytical calculations}

We apply the same "weakly incoherent"\cite{MosesMcKenzie1999} model as in
Refs. \cite{WIPRB2011,WIFNT2011,WIJETP2011}, i.e. we start from isolated 2D
metallic layers with disorder, taken into account, at least, in the
self-consistent Born approximation, and consider the interlayer tunneling as
a weakest perturbation in the minimal non-vanishing order. The interlayer
conductivity is calculated using the Kubo formula\cite{Mahan} in the second
order in the interlayer tunneling $t_{z}$, taking into account only two
adjacent conducting layers. As was shown in Ref. \cite{GrigPRB2013}, this
approach is valid at $t_{z}\ll \Gamma ,\hbar \omega _{c}$. The positions of
short-range impurities on adjacent layers are assumed to be uncorrelated,
which allows the independent averaging over disorder for each conducting
layer. Then the interlayer conductivity $\sigma _{zz}\left( B_{z}\right) $
is expressed via the disorder-averaged electron Green's functions $%
\left\langle G_{R}({\boldsymbol{r}},j,\varepsilon )\right\rangle
=\left\langle G_{R}({\boldsymbol{r}}_{1}-{\boldsymbol{r}}_{2},j,\varepsilon
)\right\rangle $ on 2D conducting layer with number $j$ (see Eq. (12) of
Ref. \cite{WIJETP2011}):%
\begin{eqnarray}
\sigma _{zz} &=&\frac{2\sigma _{0}\Gamma _{0}}{\pi \nu _{2D}}\int d^{2}{%
\boldsymbol{r}}\int d\varepsilon \left[ -n_{F}^{\prime }(\varepsilon )\right]
\label{KuboB} \\
&&\times \left\langle \text{Im}G_{R}({\boldsymbol{r}},j,\varepsilon
)\right\rangle \left\langle \text{Im}G_{R}({\boldsymbol{r}},j+1,\varepsilon
)\right\rangle ,  \notag
\end{eqnarray}%
where $n_{F}^{\prime }(\varepsilon )=-1/\{4T\cosh ^{2}\left[ (\varepsilon
-\mu )/2T\right] \}$ is the derivative of the Fermi distribution function, $%
\mu $ is the chemical potential,
\begin{equation}
\sigma _{0}=e^{2}t_{z}^{2}\nu _{2D}d/\hbar \Gamma _{0}  \label{sig0}
\end{equation}%
is the interlayer conductivity without magnetic field, $\nu
_{2D}=2g_{LL}/\hbar \omega _{c}=m^{\ast }/\pi \hbar ^{2}=\nu _{3D}d$ is the
2D density of states (DoS) at the Fermi level in the absence of magnetic
field per two spin components, $d$ is the interlayer distance, and $%
g_{LL}=eB_{z}/2\pi \hbar c$ is the LL degeneracy per unit area.

The 2D metallic electron system in a perpendicular magnetic field in the
point-like impurity potential has been extensively studied.\cite%
{Ando1,Fogler1997,KukushkinUFN1988,DmitrievRMP2012,QHE,HuckesteinRMP1995,Ando2,Ando3,Baskin,Brezin,Marihin1989,Imp,Burmi}
In the self-consistent single-site approximation\cite{Ando2}, which takes
into account all diagrams without intersection of impurity lines,\cite%
{CommentCrossing} the coordinate electron Green's function, averaged over
impurity configurations, is given by%
\begin{equation}
G({\boldsymbol{r}}_{1},{\boldsymbol{r}}_{2},\varepsilon )=\sum_{n,k_{y}}\Psi
_{n,k_{y}}^{0\ast }(r_{2})\Psi _{n,k_{y}}^{0}(r_{1})G\left( \varepsilon
,n\right) ,  \label{Gg}
\end{equation}%
where $\Psi _{n,k_{y}}^{0}(r_{1})$ are the 2D electron wave functions in a
perpendicular magnetic field,\cite{LL3} and the 2D electron Green's function
$G\left( \varepsilon ,n\right) $ does not depend on $k_{y}$:%
\begin{equation}
G\left( \varepsilon ,n\right) =\frac{1}{\varepsilon -\hbar \omega _{c}\left(
n+1/2\right) -\Sigma \left( \varepsilon \right) },  \label{Gn}
\end{equation}%
where we have used that the 2D electron dispersion in magnetic field $%
\varepsilon _{2D}\left( n\right) =\hbar \omega _{c}\left( n+1/2\right) $,
and $\Sigma \left( \varepsilon \right) $ is the electron self-energy part
due to the scattering by impurities.

In a perpendicular-to-layers magnetic field the integration over coordinate
in Eq. (\ref{KuboB}) with the Green's functions (\ref{Gg}) reduces to the
normalization of the wave functions and gives (see Eq. (14) of Ref. \cite%
{WIJETP2011})
\begin{equation}
\sigma _{zz}=\frac{\sigma _{0}\Gamma _{0}\hbar \omega _{c}}{\pi }\int
d\varepsilon \left[ -n_{F}^{\prime }(\varepsilon )\right] \sum_{n}\left\vert
\text{Im}G(\varepsilon ,n)\right\vert ^{2}  \label{sp1}
\end{equation}%
with Im$G(\varepsilon ,n)$ given by Eq. (\ref{Gn}). After substitution of
Eq. (\ref{Gn}) to Eq. (\ref{sp1}), and introducing the notations
\begin{equation}
\alpha \equiv 2\pi \left( \varepsilon -\text{Re}\Sigma \left( \varepsilon
\right) \right) /\hbar \omega _{c},~\gamma \equiv 2\pi \left\vert \text{Im}%
\Sigma \left( \varepsilon \right) \right\vert /\hbar \omega _{c},
\label{notations}
\end{equation}
the sum over $n$ in Eq. (\ref{sp1}) gives%
\begin{eqnarray}
\frac{\sigma _{zz}}{\sigma _{0}} &=&\int d\varepsilon \frac{-n_{F}^{\prime
}(\varepsilon )\Gamma _{0}}{\left\vert \text{Im}\Sigma \left( \varepsilon
\right) \right\vert }\left[ \frac{\left\vert \sinh \gamma \right\vert }{%
\cosh \gamma +\cos \alpha }\right.  \notag \\
&&\left. -\gamma \frac{\cos \alpha \,\cosh \gamma +1}{\left[ \cosh \gamma
+\cos \alpha \right] ^{2}}\right]  \label{szz}
\end{eqnarray}%
in agreement with Eqs. (19)-(21) of Ref. \cite{ChampelMineev} or with Eq.
(C3) of Ref. \cite{GM1}.

The expressions for interlayer conductivity $\sigma _{zz}$ contain the
electron self-energy $\Sigma \left( \varepsilon \right) $ coming from the
scattering on impurity potential $V_{i}\left( \mathbf{r}\right) $. The
impurities are assumed to be short-range (point-like) and randomly
distributed with volume concentration $n_{i}$:%
\begin{equation}
V_{i}\left( \mathbf{r}\right) =U\sum_{i}\delta ^{3}\left( \mathbf{r}-\mathbf{%
r}_{i}\right) .  \label{Vi}
\end{equation}%
The scattering by impurity potential given by Eq. (\ref{Vi}) is
spin-independent. In the self-consistent single-site (non-crossing)
approximation the electron self energy satisfies the following equation:\cite%
{Ando2}
\begin{equation}
\Sigma (\varepsilon )=\frac{n_{i}U}{1-UG\left( \varepsilon \right) },
\label{SER}
\end{equation}%
where the Green's function
\begin{eqnarray}
G\left( \varepsilon \right) &=&\sum_{n,k_{y},k_{z}}G\left( \varepsilon
,n\right) =\frac{g_{LL}}{d}\sum_{n}G\left( \varepsilon ,n\right)  \label{Gr0}
\\
&=&-\frac{\pi g_{LL}}{\hbar \omega _{c}d}\tan \left[ \pi \frac{\varepsilon
-\Sigma (\varepsilon )}{\hbar \omega _{c}}\right] .  \label{Gr}
\end{eqnarray}%
The summation over $k_{y}$ in Eq. (\ref{Gr0}) gives the LL degeneracy $%
g_{LL} $, and the summation over $k_{z}$ gives $1/d$. It is convenient to
use the normalized electron Green's function
\begin{equation}
g\left( \varepsilon \right) \equiv G\left( \varepsilon \right) \,\hbar
\omega _{c}d/\pi g_{LL}.  \label{gnorm}
\end{equation}

To obtain the monotonic growth of longitudinal interlayer magnetoresistance,
the self-consistent Born approximation (SCBA) is sufficient, which gives
instead of Eq. (\ref{SER})%
\begin{equation}
\Sigma (\varepsilon )-n_{i}U=n_{i}U^{2}G\left( \varepsilon \right) =\Gamma
_{0}g\left( \varepsilon \right) .  \label{SigSCBA}
\end{equation}%
Here we used that the zero-field level broadening is $\Gamma _{0}=\pi
n_{i}U^{2}\nu _{3D}=\pi n_{i}U^{2}g_{LL}/d\hbar \omega _{c}$. Below we also
neglect the constant energy shift $n_{i}U$ in Eq. (\ref{SigSCBA}), which
does not affect physical quantities as conductivity.

Eqs. (\ref{Gr})-(\ref{SigSCBA}) give the following equations on $g\equiv
g\left( \varepsilon \right) $:%
\begin{equation}
\text{Im}g=\frac{\sinh \left( 2\pi \Gamma _{0}\text{Im}g/\hbar \omega
_{c}\right) }{\cosh \left( 2\pi \Gamma _{0}\text{Im}g/\hbar \omega
_{c}\right) +\cos \left( 2\pi \varepsilon ^{\ast }/\hbar \omega _{c}\right) }%
,  \label{Img}
\end{equation}%
\begin{equation}
\text{Re}g=\frac{-\sin \left( 2\pi \varepsilon ^{\ast }/\hbar \omega
_{c}\right) }{\cosh \left( 2\pi \Gamma _{0}\text{Im}g/\hbar \omega
_{c}\right) +\cos \left( 2\pi \varepsilon ^{\ast }/\hbar \omega _{c}\right) }%
.  \label{Reg}
\end{equation}%
where
\begin{equation}
\varepsilon ^{\ast }\equiv \varepsilon -\text{Re}\Sigma ^{R}(\varepsilon
)=\varepsilon -\Gamma _{0}\text{Re}g(\varepsilon ).  \label{estar}
\end{equation}

These equations can be written also for $\Sigma ^{R}(\varepsilon )$. With
notations $\gamma _{0}=2\pi \Gamma _{0}/\hbar \omega _{c}$, $\gamma \equiv
2\pi \text{Im}\Sigma ^{R}(\varepsilon )/\hbar \omega _{c}$, $\alpha \equiv
2\pi \varepsilon ^{\ast }/\hbar \omega _{c}$, $\delta \equiv -2\pi $Re$%
\Sigma ^{R}(\varepsilon )/\hbar \omega _{c}=2\pi \left( \varepsilon ^{\ast
}-\varepsilon \right) /\hbar \omega _{c}$, Eqs. (\ref{SigSCBA}) and (\ref{Gr}%
) give
\begin{equation}
\frac{\gamma }{\gamma _{0}}=\frac{\sinh \left( \gamma \right) }{\cosh \left(
\gamma \right) +\cos \left( \alpha \right) },  \label{gamma}
\end{equation}%
\begin{equation}
\delta \equiv \alpha -\frac{2\pi \varepsilon }{\hbar \omega _{c}}=\frac{%
\gamma _{0}\sin \left( \alpha \right) }{\cosh \left( \gamma \right) +\cos
\left( \alpha \right) }.  \label{alpha}
\end{equation}%
The solution of Eq. (\ref{gamma}) gives Im$\Sigma (\alpha )$, while Eq. (\ref%
{alpha}) allows to find $\alpha \left( \varepsilon \right) $ and Re$\Sigma
(\varepsilon )$. The system of Eqs. (\ref{gamma}) and (\ref{alpha}) differs
from Eq. (30) of Ref. \cite{ChampelMineev} even in the absence of electron
reservoir (at $R=0$), because in Eq. (30) of Ref. \cite{ChampelMineev} the
oscillating real part of the electron self energy is neglected, which leads
to a different dependence of $\sigma _{zz}(B_{z})$. Eqs. (\ref{szz}),(\ref%
{gamma}) and (\ref{alpha}) will be used for numerical calculations in the
next section.

Eq. (\ref{gamma}) allows to find the value $\gamma _{0c}$, when
the LLs become isolated in SCBA, i.e. when the DoS and Im$\Sigma
^{R}(\varepsilon )$ between LLs become zero. In the middle between
two adjacent LLs $\cos \left( \alpha \right) =1$,
and equation (\ref{gamma}) for $\gamma $ becomes%
\begin{equation}
\frac{\gamma }{\gamma _{0}}=\frac{\sinh \left( \gamma \right) }{\cosh \left(
\gamma \right) +1}=\tanh \left( \gamma /2\right) .  \label{EqGm}
\end{equation}%
This equation always has a trivial solution $\gamma =0$. However,
at $\gamma _{0}>\gamma _{0c}=2$, corresponding to $\pi \Gamma
_{0}>\hbar \omega _{c}$, Eq. (\ref{EqGm}) also has a non-zero
solution. This nonzero solution means a finite DoS at energy
between LLs. In the next section we obtain that this crossover at
$\hbar \omega _{c}=\pi \Gamma _{0}$ also affects the monotonic
part of interlayer magnetoresistance.

\subsection{High-field limit}

In the high-field limit, the monotonic growth of longitudinal interlayer MR $%
R_{zz}\left( B_{z}\right) $, given by Eq. (\ref{Rzz0}), was calculated for
the Lorentzian LL shape in Refs. \cite{WIPRB2011,WIFNT2011}. In Ref. \cite%
{WIJETP2011} $R_{zz}\left( B_{z}\right) $ was calculated in the non-crossing
approximation, but the coefficient $\eta $ in Eq. (24) of Ref. \cite%
{WIJETP2011} is greater than the correct value by a factor $4/3$.\cite%
{CommentNC} Following the procedure of Ref. \cite{WIJETP2011}, we calculate $%
R_{zz}\left( B_{z}\right) $ in the SCBA at $\hbar \omega _{c}\gg \Gamma _{0}$
to compare with the numerical results in Sec. III. At $\hbar \omega _{c}\gg
\Gamma _{0}$ the summation over $n$ in Eq. (\ref{Gr0}) restricts to only one
LL $n=n_{F}$ on the Fermi level and gives the equation for $G\left( \Delta
\varepsilon \right) =\left( g_{LL}/d\right) G\left( \varepsilon
,n_{F}\right) $:
\begin{equation}
G\left( \Delta \varepsilon \right) =g_{LL}/d/\left[ \Delta \varepsilon
-n_{i}U^{2}G\left( \Delta \varepsilon \right) \right] ,  \label{G1}
\end{equation}%
where we have used Eq. (\ref{SigSCBA}) and the notation $\Delta \varepsilon
=\varepsilon -\hbar \omega _{c}\left( n_{F}+1/2\right) -n_{i}U$. This
equation yields%
\begin{equation}
\text{Im}G\left( \Delta \varepsilon ,n_{F}\right) =\frac{\pi }{2\Gamma
_{0}\hbar \omega _{c}}\sqrt{4\hbar \omega _{c}\Gamma _{0}/\pi -\left( \Delta
\varepsilon \right) ^{2}},  \label{ImG}
\end{equation}%
which is nonzero only at $\left\vert \Delta \varepsilon \right\vert <\Gamma
_{B}\equiv \sqrt{4\hbar \omega _{c}\Gamma _{0}/\pi }$. Substituting Eq. (\ref%
{ImG}) to Eq. (\ref{sp1}), keeping only one LL at the Fermi level and
averaging over MQO period, we get%
\begin{eqnarray}
\sigma _{zz} &=&\frac{\sigma _{0}\Gamma _{0}}{\pi }\left( \frac{\pi }{%
2\Gamma _{0}\hbar \omega _{c}}\right) ^{2}\int_{-E_{1}}^{E_{1}}d\varepsilon %
\left[ \Gamma _{B}^{2}-\left( \Delta \varepsilon \right) ^{2}\right]  \notag
\\
&=&\frac{\sigma _{0}\Gamma _{0}}{\pi }\left( \frac{\pi }{2\Gamma _{0}\hbar
\omega _{c}}\right) ^{2}\frac{4\Gamma _{B}^{3}}{3}=\frac{8\sigma _{0}}{3%
\sqrt{\pi }}\sqrt{\frac{\Gamma _{0}}{\hbar \omega _{c}}},  \label{sLim}
\end{eqnarray}%
corresponding to $\eta =3\sqrt{\pi }/8$ in Eq. (\ref{Rzz0}). The SCBA and
noncrossing approximation coincide at $c_{i}\gg 1$ and, therefore, give the
same value $\eta =3\sqrt{\pi }/8$.

\section{Numerical results and discussion}

\begin{figure}[tb]
\includegraphics[width=0.49\textwidth]{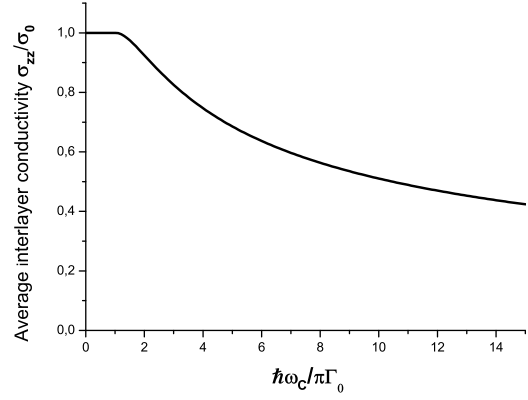}
\caption{Average interlayer conductivity $\bar{\protect\sigma}_{zz}$ as
function of the LL separation $\hbar \protect\omega _{c}/\protect\pi \Gamma
_{0}\propto B_{z}$, calculated numerically using Eqs. (\protect\ref{gamma}),
(\protect\ref{alpha}) and (\protect\ref{szz}).}
\label{Fig1}
\end{figure}
\begin{figure}[tb]
\includegraphics[width=0.49\textwidth]{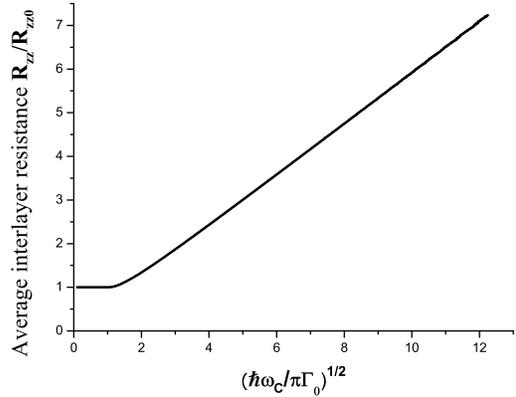}
\caption{Average interlayer resistance $\bar{R}_{zz}=1/\bar{\protect\sigma}%
_{zz}$ as function of a square root of $\protect\sqrt{\hbar \protect\omega %
_{c}/\protect\pi \Gamma _{0}}\propto \protect\sqrt{B_{z}}$, calculated
numerically using Eqs. (\protect\ref{gamma}), (\protect\ref{alpha}) and (%
\protect\ref{szz}). At low field $\hbar \protect\omega _{c}<\protect\pi %
\Gamma _{0}$, $\bar{R}_{zz}\approx const$.}
\label{Fig2}
\end{figure}
\begin{figure}[tb]
\includegraphics[width=0.49\textwidth]{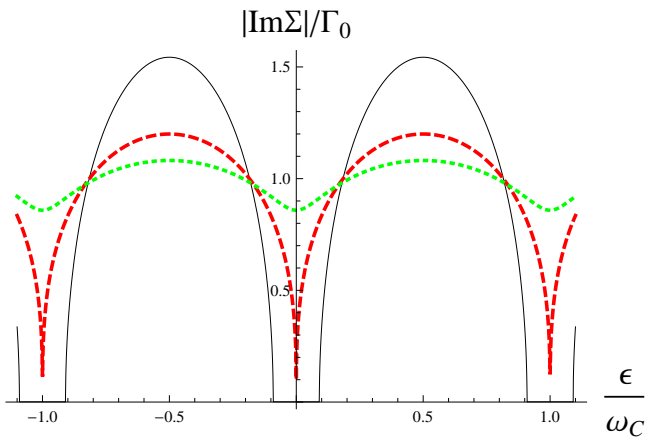}
\caption{(Color online) Imaginary part of the electron self energy Im$\Sigma
/\Gamma _{0}$ as function of energy $\epsilon$ calculated from Eqs. (\protect\ref{gamma}%
) and (\protect\ref{alpha}) for three different values of $\hbar \protect%
\omega _{c}/\Gamma _{0}=2\protect\pi $ (solid black line), $\hbar \protect%
\omega _{c}/\Gamma _{0}=\protect\pi $ (dashed red line), and $\hbar \protect%
\omega _{c}/\Gamma _{0}=2\protect\pi /3$ (dotted green line). When Im$\Sigma
=0$, the DoS is also zero. The critical field $B_{c}$, when the LL become
first separated, corresponds to $\hbar \protect\omega _{c}/\Gamma _{0}=%
\protect\pi $.}
\label{Fig3}
\end{figure}

Substituting the solutions of Eqs. (\ref{gamma}) and (\ref{alpha}) into Eq. (%
\ref{szz}) one can calculate interlayer conductivity $\sigma _{zz}$
numerically in the SCBA in the full interval of magnetic field. The result
is shown in Figs. \ref{Fig1} and \ref{Fig2}. As one can see from Fig. \ref%
{Fig2}, in high field the calculated dependence $R_{zz}\left( B_{z}\right)
\approx \left( 3\sqrt{\pi }/8\right) \sqrt{\hbar \omega _{c}/\Gamma _{0}}%
\propto B_{z}^{1/2}$ in agreement with Eq. (\ref{sLim}) and Refs. \cite%
{WIPRB2011,WIFNT2011,WIJETP2011,GrigPRB2013}. From Fig. \ref{Fig1} one can
clearly see, that the drop of interlayer conductivity $\sigma _{zz}\left(
B_{z}\right) $ starts not from zero field, but from some critical field $%
B_{c}$, where $\hbar \omega _{c}=\pi \Gamma _{0}$, in agreement with the
analytically obtained crossover at $\gamma _{0c}$ of the solution of Eq. (%
\ref{EqGm}). At this field in SCBA the Landau levels become isolated, i.e.
the LL separation $\hbar \omega _{c}$ exceeds the LL broadening $2\Gamma $
(see Fig. \ref{Fig3}). Below this field, at $B<B_{c}$, $\sigma _{zz}\left(
B_{z}\right) $ is flat within the accuracy of our calculation. This means,
that the field dependence of the monotonic part of longitudinal MR $\bar{R}%
_{zz}\left( B_{z}\right) $ is not a simple analytic function, as was assumed
in Refs. \cite{WIPRB2011,WIPRB2012} [see Eq. (\ref{Rzz1})]: within SCBA it
is constant at $B<B_{c}$ and starts to grow at $B>B_{c}$, reaching the
dependence $R_{zz}\left( B_{z}\right) \propto B_{z}^{1/2}$ at $\hbar \omega
_{c}\gg \Gamma _{0}$. Such crossover from low-field flat to the high-field
increasing MR $\bar{R}_{zz}\left( B_{z}\right) $ was observed in the
strongly anisotropic quasi-2D organic metal $\beta ^{\prime \prime }$-
(BEDT-TTF)$_{2}$SF$_{5}$CH$_{2}$CF$_{2}$SO$_{3}$ at $B\approx 8$T.\cite{W2}

The predicted crossover of MR at $B=B_{c}$ needs further
theoretical investigation. The SCBA assumes sharp edges of the
electron DoS for each LL in strong magnetic field. It works well
as a zero approximation, capturing rough physical effects, such as
the monotonic growth of MR $R_{zz}\left( B_{z}\right) \propto
B_{z}^{1/2}$ in strong field. However, more elaborated theories
predict
exponential tails of the electronic DoS for each LL,\cite%
{Ando1,Fogler1997,KukushkinUFN1988,DmitrievRMP2012,QHE,HuckesteinRMP1995,Ando2,Ando3,Baskin,Brezin,Marihin1989,Imp,Burmi}
which may lead to the small deviations from the flat average MR $\bar{R}%
_{zz}\left( B_{z}\right) $ at $B<B_{c}$.

\medskip

The work was supported by the Russian Foundation for Basic Research Grant
No. 13-02-00178.

\end{document}